\documentclass[12pt]{article}
\usepackage{comment}
\usepackage{amsmath}
\usepackage{amssymb}
\usepackage{graphicx}
\usepackage{here}
\usepackage{subcaption}
\usepackage{url}
\usepackage[sort&compress, numbers, merge]{natbib}
\usepackage{braket}
\usepackage{physics}
\usepackage[compat=1.1.0]{tikz-feynhand}
\usepackage{slashed}
\usepackage{mathtools}

\usepackage[colorlinks=true, linkcolor=blue, citecolor=blue,urlcolor=black]{hyperref}
\usepackage{cleveref}

\crefname{figure}{Fig.}{Figs.}
\crefname{equation}{Eq.}{Eqs.}

\usepackage{todonotes}

\numberwithin{equation}{section}

\usepackage[colorlinks=true, linkcolor=blue, citecolor=blue,
urlcolor=black]{hyperref}

\begin{document}
\def\ps{\mathbf{p}}
\def\PS{\mathbf{P}}
\baselineskip 0.6cm
\def\simgt{\mathrel{\lower2.5pt\vbox{\lineskip=0pt\baselineskip=0pt
           \hbox{$>$}\hbox{$\sim$}}}}
\def\simlt{\mathrel{\lower2.5pt\vbox{\lineskip=0pt\baselineskip=0pt
           \hbox{$<$}\hbox{$\sim$}}}}
\def\simprop{\mathrel{\lower3.0pt\vbox{\lineskip=1.0pt\baselineskip=0pt
             \hbox{$\propto$}\hbox{$\sim$}}}}
\def\tr{\mathop{\rm tr}}
\def\SU{\mathop{\rm SU}}

\begin{titlepage}

\begin{flushright}
IPMU23-0038
\end{flushright}

\vskip 1.1cm

\begin{center}

{\Large \bf 
Indirect Probe of Electroweak-Interacting Particles at $\mu$TRISTAN $\mu^+\mu^+$ Collider}

\vskip 1.2cm

Risshin Okabe and Satoshi Shirai
\vskip 0.5cm

{\it

{Kavli Institute for the Physics and Mathematics of the Universe
 (WPI), \\The University of Tokyo Institutes for Advanced Study, \\ The
 University of Tokyo, Kashiwa 277-8583, Japan}

}

\vskip 1.0cm

\abstract{
Recently, a novel collider, called  $\mu$TRISTAN, has been proposed, offering the capability to achieve high-energy collisions of anti-muons. 
This high-energy collider presents an exceptional opportunity for the discovery of electroweak-interacting massive particles (EWIMPs), which are predicted by various new physics models.
In a lepton collider like $\mu$TRISTAN, the potential for discovering EWIMPs extends beyond their direct production. Quantum corrections arising from EWIMP loops can significantly enhance our prospects for discovery by precise measurement of Standard Model processes.
This study focuses on the indirect detection method within the $\mu$TRISTAN experiment, with a specific emphasis on TeV-scale EWIMP dark matter scenarios that yield the correct thermal relic density. At collision energies for $ \sqrt{s} = O(1-10)$ TeV, these EWIMPs introduce noticeable effects, typically in the range of $O(0.1-1)$\%.
Our findings indicate that at $\sqrt{s} = 2\, (10)$ TeV, with an integrated luminosity of 10 ab$^{-1}$, the $\mu$TRISTAN can detect Higgsino at a mass of 1.3 (3.0) TeV and Wino at a mass of 1.9 (4.4) TeV, assuming an optimistic level of systematic uncertainty in the observation of the Standard Model processes. 
}

\end{center}
\end{titlepage}

%-----------------------------------------------------
\section{Introduction} 
%-----------------------------------------------------

The Standard Model (SM) of particle physics represents a remarkable milestone in our understanding of the fundamental laws governing the Universe. It offers an elegant and precise description of numerous experimental and observational findings.
Yet, the SM is far from perfect, as it grapples with several unsolved mysteries, one of which is the presence of Dark Matter (DM).

To address such problems, we have been working on extensions to the SM that introduce new particles. Some of models beyond the SM (BSM) include electroweak-interacting massive particles (EWIMPs), and they can be potential solutions to the dark matter puzzle.
One intriguing idea is that DM itself can be an EWIMP, which links it more closely to the electroweak forces described by the SM.
The electroweak interaction plays a pivotal role in facilitating the freeze-out mechanism of DM to explain the correct relic abundance. 
This electroweak charge is instrumental in enhancing the stability of DM. 
Furthermore, the electroweak interaction holds a central position in the pursuit of DM detection, spanning direct/indirect detection methods and collider experiments.

These theoretical extensions, such as supersymmetry (SUSY), suggest the existence of DM particles with electroweak charges, such as Wino, Higgsino, and the Minimal Dark Matter (MDM) model \cite{Cirelli:2005uq, *Cirelli:2007xd, *Cirelli:2009uv}.
While collider experiments are essential for exploring DM, it is not easy to find electroweak-charged DM. The DM does not interact directly with detectors, making it a difficult target to observe.

One important strategy is to consider $\mathrm{SU}(2)_L$ isospin partners, which are particles closely related to DM. These partners have nearly the same mass as DM and can possess a macroscopic lifetime \cite{Chen:1996ap,Thomas:1998wy, Nagata:2014wma, Ibe:2022lkl}.
These meta-stable isospin partners produce several exotic tracks at collider experiments.
The search for disappearing charged tracks from the charged isospin partners places significant constraints on the EWIMPs \cite{Ibe:2006de, *Buckley:2009kv, Asai:2007sw, *Asai:2008sk, *Asai:2008im,Mahbubani:2017gjh,Fukuda:2017jmk}. 
The decay products of the isospin partner can generate displaced and soft tracks, each yielding distinct signatures that markedly diverge from those of SM events \cite{Fukuda:2019kbp}.

However, the search based on the direct EWIMP production is strongly dependent on the details of the EWIMP decay: for example, in the search for disappearing charged tracks, the sensitivity is strongly dependent on the lifetime of the EWIMP's charged partner. The energies of the decay products such as leptons also depend on the mass splitting of the EWIMP, which strongly affects the EWIMP search~\cite{Ibe:2023dcu}. For instance, in the case of the pure Higgsino, which has a mass splitting of about 350 MeV, the search at the Large Hadron Collider (LHC) by disappearing charged tracks is the most sensitive, setting a lower mass limit of approximately 200 GeV~\cite{ATLAS:2022rme}. However, when the mass splitting is roughly 1 GeV, the LHC search is very challenging, and there is no established limit by the LHC.

Another approach involves indirect searches that rely on the effects of EWIMPs loops that can influence the behavior of SM processes observed in collider experiments.
At $e^+ e^- $colliders like the International Linear Collider, the influence of EWIMP loops is indirect but profound, impacting the precision measurement of di-fermion $e^+ e^- \to f \bar{f}$ production \cite{Harigaya:2015yaa}. 
The similar process can also be utilized in the $\mu^+ \mu^-$ collider~\cite{Franceschini:2022sxc}.
This effect offers us a powerful way to detect EWIMPs, sometimes even outperforming the detection capabilities of direct EWIMP production.
Similarly, at hadron colliders, the EWIMPs have an indirect impact on the Drell-Yan process, making it a valuable tool for detecting EWIMPs \cite{Matsumoto:2017vfu,Chigusa:2018vxz,DiLuzio:2018jwd,Matsumoto:2018ioi,Katayose:2020one}.
This approach barely depends on the properties of the EWIMP decay, such as the lifetime of charged partners.
Thus the indirect search provides a generic prospect for the EWIMP search.

A recent development is the proposal for a new lepton collider $\mu$TRISTAN: the $e^-\mu^+$ and $\mu^+\mu^+$ collider \cite{Hamada:2022mua}.
This collider utilizes a low-emittance muon beam, originally developed for the muon g-2/EDM experiment at J-PARC \cite{Abe:2019thb}. 
This beam enables high-energy $\mu^+\mu^+$  scattering, which has inspired the concept of this new collider. 
Motivated by this new proposed collider, we will explore the possibilities and potential for the EWIMP detection.
Unlike conventional lepton colliders, this innovative collider concept presents specific challenges due to the limited production of DM. This is because we require multi-body processes, such as $\mu^+ \mu^+ \to \mu^+ \mu^+ + \mathrm{EWIMP} + \mathrm{EWIMP}$ via the vector boson fusion for the EWIMPs production, which not only reduces the production cross section but also adds complexity to the analysis \cite{Kitano:2023pascos}.
One of the advantages of using lepton colliders is that they are excellent at making very accurate measurements, and these precise measurements are a valuable tool in our search for BSM \cite{Hamada:2022uyn}.

In this context, we present a novel method that exploits the loop effects from the BSM to address the EWIMPs in the unique environment of the $\mu^+\mu^+$ collider. 
In this paper, we investigate the effect of EWIMP loops on the M{\o}ller-like scattering, as shown in Fig.\,\ref{fig:diagram}. 
Since the beam energy and luminosity of the $\mu^+\mu^+$ collider are currently still under consideration, the purpose of this paper is to identify which masses of EWIMPs can be searched indirectly with a given collider configuration.
As we will discuss later, we find that TeV-scale EWIMPs can provide an $O(0.1-1)$\% correction to these processes.
By these indirect methods, we can detect TeV-scale EWIMPs. These findings will be important in the search for TeV-scale DM physics.

%-----------------------------------------------------
\section{Indirect Signals from EWIMPs}
%-----------------------------------------------------

\begin{figure}[t]
    \centering
    \includegraphics[width=0.3\textwidth]{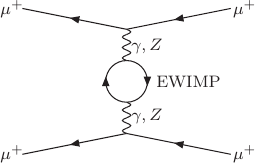}
    \caption{An example of the diagrams of the loop correction from the EWIMPs for the process $\mu^+\mu^+ \to \mu^+\mu^+$.}
    \label{fig:diagram}
\end{figure}

We investigate the impact of EWIMPs on $\mu^+\mu^+$ elastic scattering. The primary contribution arises from the one-loop process illustrated in Fig.\,\ref{fig:diagram}. 
Integrating out the EWIMP with a mass $m$ leads to an effective theory describing the correction to the SM process by the EWIMP. At the one-loop level, the effective Lagrangian reads
\begin{align}
    \mathcal{L}_\mathrm{eff} = \mathcal{L}_\mathrm{SM} + \frac{g^2 C_{WW}}{8} W^a_{\mu\nu} \Pi_\mathrm{BSM} (-D^2/m^2) W^{a\mu\nu} + \frac{g^{\prime 2} C_{BB}}{8} B_{\mu\nu} \Pi_\mathrm{BSM}(-\partial^2/m^2) B^{\mu\nu} + \cdots\, ,
    \label{eq:effective}
\end{align}
where $\mathcal{L}_\mathrm{SM}$ is the SM Lagrangian, and $g$ ($g'$) is the gauge coupling of $\SU(2)_L$ ($\mathrm{U}(1)_Y$) whose field strength is denoted by $W^a_{\mu\nu}$ ($B^{\mu\nu}$) with $D^\mu$ being the covariant derivative on $W^a_{\mu\nu}$.
The ellipsis denotes operators composed of field strength tensors exceeding two; however, these are irrelevant to the subsequent discussion.
The coefficients $C_{WW}$ and $C_{BB}$ are given by
\begin{align}
    C_{WW} &= \frac{n(n-1)(n-2)}{6} \left\{
        \begin{aligned}
            &1 && \text{(Complex scalar)} \\
            &8 && \text{(Dirac fermion)}
        \end{aligned}
    \right. , \\
    C_{BB} &= 2nY^2 \left\{
        \begin{aligned}
            &1 && \text{(Complex scalar)} \\
            &8 && \text{(Dirac fermion)}
        \end{aligned}
    \right. .
\end{align}
We need to multiply by an additional factor of $1/2$ for the case of a real scalar or a Majorana fermion.
The function $\Pi_\mathrm{BSM}(x)$ represents the self-energy of the gauge bosons resulting from the EWIMP loop. Employing the $\overline{\text{MS}}$ regularization scheme with the renormalization scale set at $\mu=m$, we can derive its explicit expression as follows:
\begin{align}
    \Pi_\mathrm{BSM}(x) = \left\{
        \begin{aligned}
            &\frac{1}{16\pi^2} \int_0^1 \mathrm{d}y\, y(1-y) \ln[1-y(1-y)x] && \text{(Fermion)} \\
            &\frac{1}{16\pi^2} \int_0^1 \mathrm{d}y\, (1-2y)^2 \ln[1-y(1-y)x] && \text{(Scalar)} \\
        \end{aligned}
    \right. .
\end{align}

As described in the introduction, the $\mu$TRISTAN experiment focuses on high-energy beam collisions. In our analysis, we typically consider scenarios where the beam energy is higher than the DM mass. Therefore, it is useful to illustrate the asymptotic behavior of the loop function $\Pi(x)$ in the limit of $x\to -\infty$:

\begin{align}
    \Pi_\mathrm{BSM}(x) \to \frac{3}{144\pi^2} \ln[-x] \left\{
        \begin{aligned}
            &1/2 && \text{(Fermion)} \\
            &1 && \text{(Scalar)} \\
        \end{aligned}
    \right. .
    \label{eq:Pi_asymp}
\end{align}
In order to keep the perturbative loop expansion valid, we need
\begin{align}
{g^2 C_{WW}} \Pi(x) \ll 1 \quad \text{and} \quad {g'^2 C_{BB}} \Pi(x) \ll 1. \label{eq:expansion}
\end{align}
Therefore, if we consider either excessively high momentum transfers or large electroweak charges of the BSM particles, the current one-loop analysis may become invalid. In subsequent analyses, we assume the BSM mass range is approximately 1-10 TeV, with the collider operating at up to 20 TeV. Under these benchmark conditions, and considering specific EWIMPs (Higgsino, Wino, 5-plet fermion, and 7-plet scalar) as examples, the criteria specified in Eq.\,\eqref{eq:expansion} are safely satisfied.

The effective Lagrangian \eqref{eq:effective} enables us to calculate the amplitude of the $\mu^+\mu^+\to\mu^+\mu^+$ scattering, and the leading-order (LO) amplitude is
\begin{align}
i\mathcal{M}_\mathrm{LO}(\mu^+_{h_1}(p_1) \mu^+_{h_2}(p_2) \to \mu^+_{h_3}(p_3) \mu^+_{h_4}(p_4) ) = &
-i[\bar{v}_{h_1}(p_1) \gamma^\mu v_{h_3}(p_3)]
\sum_{V = \gamma,\,Z}
\frac{C_{h_1 V}\,C_{h_2 V}}{t - m^2_V }
[\bar{v}_{h_2}(p_2) \gamma_\mu v_{h_4}(p_4)] \notag \\
&+ i[\bar{v}_{h_1}(p_1) \gamma^\mu v_{h_4}(p_4)]
\sum_{V = \gamma,\,Z}
\frac{C_{h_1 V}\,C_{h_2 V}}{u - m^2_V }
[\bar{v}_{h_2}(p_2) \gamma_\mu v_{h_3}(p_3)]\, ,
\label{eq: LO1}
\end{align}
where $v_{h_i}(p_i)$ is the Dirac-spinor wave function for the $i$-th anti-muon with the helicity $h_i=L,R$, and $t=(p_1-p_3)^2$ and $u=(p_1-p_4)^2$ are the Mandelstam variables. The couplings $C_{h_i V}$ are 
\begin{align}
    C_{L Z} &= -g_Z s_W^2\, , \\
    C_{R Z} &= g_Z \left(\frac{1}{2} - s_W^2\right)\, , \\
    C_{L \gamma} &= C_{R \gamma} = +e\, ,
\end{align}
where $g_Z = g/c_W$ with $s_W$ ($c_W$) being the sine (cosine) of the Weinberg angle $\theta_W$.

The next-to-leading-order (NLO) contribution by EWIMPs, can be computed as  
\begin{align}
    & i\mathcal{M}_\mathrm{BSM}(\mu^+_{h_1} \mu^+_{h_2} \to \mu^+_{h_3} \mu^+_{h_4}) 
    \notag \\
    & \ \ = 
-i[\bar{v}_{h_1}(p_1) \gamma^\mu v_{h_3}(p_3)]
\sum_{V\,V' = \gamma,\,Z}
\frac{C_{h_1 V}\,C_{h_2 V'}\,d_{VV'}\,t\,\Pi_\mathrm{BSM}(t/m^2)
}
{(t - m^2_V ) (t - m^2_{V'})}
[\bar{v}_{h_2}(p_2) \gamma_\mu v_{h_4}(p_4)] \notag \\
&\ \ \ \  +i[\bar{v}_{h_1}(p_1) \gamma^\mu v_{h_4}(p_4)]
\sum_{V\,V' = \gamma,\,Z}
\frac{C_{h_1 V}\,C_{h_2 V'}\,d_{VV'}\,u\,\Pi_\mathrm{BSM}(u/m^2)
}
{(u - m^2_V ) (u - m^2_{V'})}
[\bar{v}_{h_2}(p_2) \gamma_\mu v_{h_3}(p_3)]\, .
\label{eq:BSM}
\end{align}
Here the gauge group factors $d_{VV'}$ are given by
\begin{align}
    d_{ZZ} &= \frac{g_Z^2}{2} (c_W^4 C_{WW} + s_W^4 C_{BB}) \, , \\
    d_{\gamma\gamma} &= \frac{e}{2} (C_{WW} + C_{BB}) \, , \\
    d_{Z\gamma} &= d_{\gamma Z} = \frac{e g_Z}{2} (c_W^2 C_{WW} - s_W^2 C_{BB}) \, .
\end{align}

Given that the BSM contribution is small, as indicated by Eq.\,\eqref{eq:expansion}, the leading BSM effect arises from the interference between the SM and BSM matrix elements.
Using the matrix elements \eqref{eq: LO1} and \eqref{eq:BSM}, we can estimate the leading contribution to the $\mu^+\mu^+ \to \mu^+\mu^+$ differential cross section from the BSM particles by
\begin{align}
    \dv{\sigma_{\mathrm{BSM}}}{\cos\theta}
        \simeq \frac{1}{32\pi s} \mathrm{Re}({\mathcal{M}_{\mathrm{LO}}  \mathcal{M}^*_{\mathrm{BSM}}})\, .
\end{align}
By using the difference between the contributions from BSM and the cross section in the SM $\dv*{\sigma_\mathrm{SM}}{\cos\theta}$, we conduct an indirect search for BSM. 
As we consider collision energies greater than the weak scale, electroweak loops within the SM also provide significant corrections.
In this paper, we will use the NLO values within the SM as the SM cross section $\sigma_{\mathrm{SM}}$.
To estimate the NLO SM cross section $\mu^+\mu^+\to \mu^+\mu^+(\gamma)$, we use a slightly modified version of the program \textsc{aITALC}\cite{Lorca:2004fg}, which integrates several tools: \textsc{Qgraf}\cite{Nogueira:1991ex}, \textsc{Diana}\cite{Tentyukov:1999is}, \textsc{Form}\cite{Vermaseren:2000nd}, \textsc{LoopTools}\cite{Hahn:1998yk}, and \textsc{FF}\cite{vanOldenborgh:1990yc}.
In this work, we have incorporated an additional cut on the photon energy, setting $E^{\max}_{\gamma} = 0.1 \sqrt{s}$, where $s=(p_1 + p_2)^2$.
In this analysis, we omit the real emission of $Z$ bosons; however, we find that this correction amounts to only a few percent and does not significantly alter the results of the present analysis.

\begin{figure}[t!]
	\centering
 	\subcaptionbox{$\sqrt{s} = 1$ TeV \label{fig:1TeV_cross}}
	{\includegraphics[ width=0.49\textwidth]{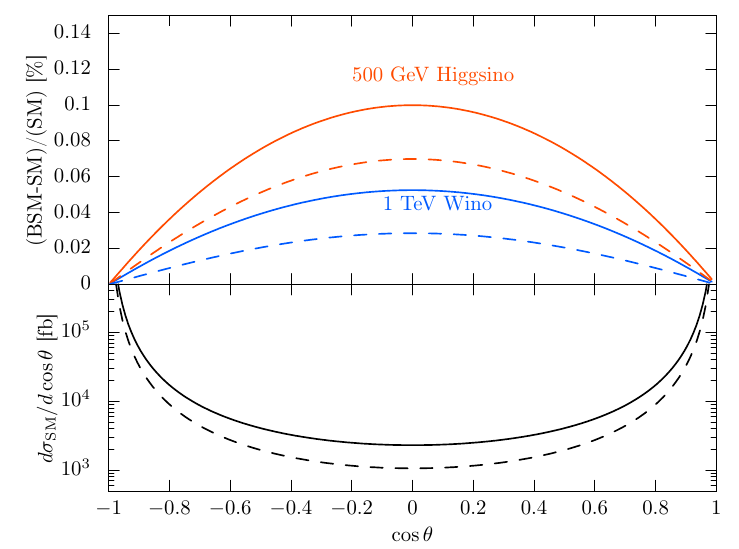}}
 	\subcaptionbox{ $\sqrt{s} = 10$ TeV  \label{fig:10TeV_cross}}
	{\includegraphics[width=0.49\textwidth]{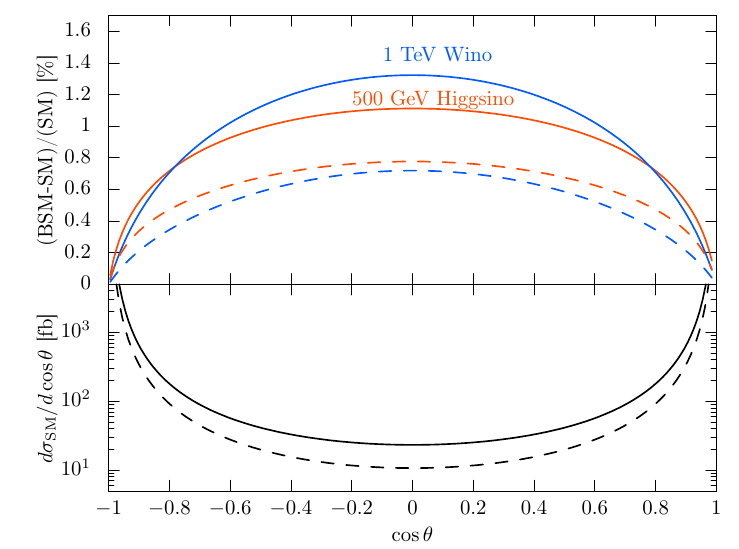}}
 \caption{
 The correction from the BSM particles and the SM cross section of $\mu^+\mu^+ \to \mu^+\mu^+$.
The solid lines represent that both initial anti-muons are right-handed ($P_{\mu^+}=+100$\%), while the dashed lines represent the case of the unpolarized initial state ($P_{\mu^+}=0$\%).
 }
 \label{fig:correction}
\end{figure}

The deviation
\begin{align}
    \dv{\sigma_{\mathrm{BSM}}}{\cos\theta} \biggm/ \dv{\sigma_\mathrm{SM}}{\cos\theta}
    \label{eq:correction}
\end{align}
 is plotted in Fig.\,\ref{fig:correction} for the case of 
1 TeV Wino (Majorana fermion with $n=3$ and $Y=0$) and 500 GeV Higgsino (Dirac fermion with $n=2$ and $Y=\pm 1/2$), together with the SM cross section.
One can find that if the initial beams are polarized such that incoming anti-muons are all right-handed, the contribution from EWIMPs is increased as expected.
Furthermore, it can also be found that the deviation becomes more significant for higher beam energy $\sqrt{s}$ and larger scattering angle $\theta$.
This is consistent with Eq.\,\eqref{eq:Pi_asymp}, which implies that the EWIMP effect grows logarithmically, like $\ln[t] = \ln[-s(1-\cos\theta)/2]$.

%-----------------------------------------------------
\section{Prospects of the Indirect EWIMP search}
%-----------------------------------------------------

\subsection{Analysis}

\begin{figure}[t!]
	\centering
 	\subcaptionbox{500 GeV Higgsino\label{fig:higgsino_chi}}
	{\includegraphics[ width=0.49\textwidth]{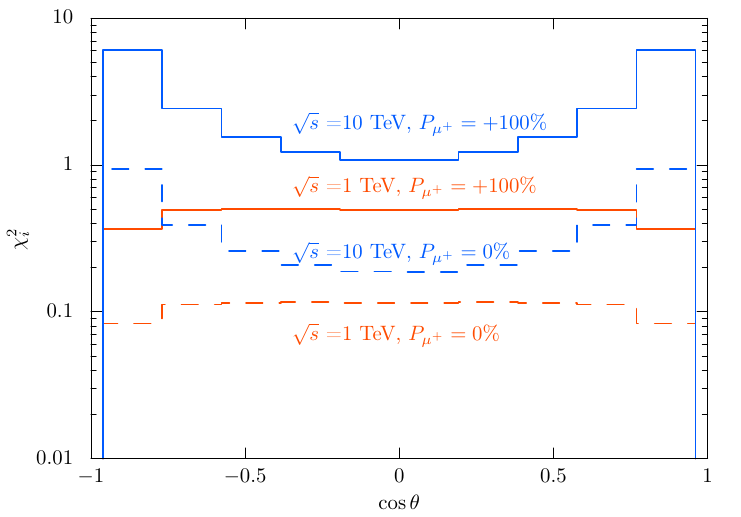}}
 	\subcaptionbox{ 1 TeV Wino \label{fig:wino_chi}}
	{\includegraphics[width=0.49\textwidth]{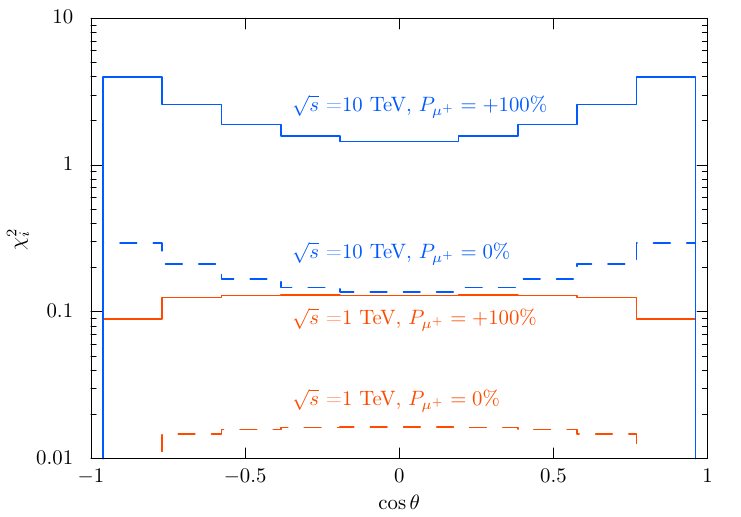}}
 \caption{
The values of $\chi_i^2$ with $\epsilon_i=0$ for each bin for an integrated luminosity $1\,\mathrm{ab}^{-1}$.
 }
 \label{fig:chi}
\end{figure}

We can estimate the sensitivity of the search for BSM by utilizing the correction to the SM cross section calculated in the previous section.
For a more statistical evaluation, we adopt the binned likelihood analysis for the differential cross section of  $\mu^+ \mu^+ \to \mu^+ \mu^+ $.
For binning, we use ten uniform intervals for $\cos\theta \in [\cos\theta_{1}:\cos\theta_{2}]$. As for the minimum and maximum angle, we use $\theta_{1} = 164^\circ$ and $\theta_{2} = 16^\circ$ as in Ref.\,\cite{Hamada:2022uyn}.
We define the following $\chi^2$ valuable,
\begin{align}
\chi^2 = \sum_{i = 1}^{10} \chi_i^2\, , ~~\mbox{where}~~
\chi_i^2 = \frac{\left[N_i^\mathrm{(BSM)}\right]^2}
{ N_i^\mathrm{(SM)} + \left[\epsilon_i\,N_i^\mathrm{(SM)}\right]^2}\, .
\label{eq: chi2}
\end{align}
Here $N_i^\mathrm{(BSM)}$ ($N_i^\mathrm{(SM)})$ is the expected value of the number of events given by
\begin{align}
    & N_i^{(\mathrm{BSM})} = L \int_{i\text{-th bin}} \mathrm{d}\cos\theta \; \left(\dv{\sigma_\mathrm{BSM}}{\cos\theta}\right)\, , \\
    & N_i^{(\mathrm{SM})} = L \int_{i\text{-th bin}} \mathrm{d}\cos\theta \; \left(\dv{\sigma_\mathrm{SM}}{\cos\theta}\right)\, ,
\end{align}
where $L$ is the integrated luminosity,
and $\epsilon_i$ in Eq.\,\eqref{eq: chi2} represents a systematic error for the $i$-th bin. 	
In Fig.\,\ref{fig:chi}, we show $\chi^2_i$ for the Higgsino and Wino with an integrated luminosity $1\,\mathrm{ab}^{-1}$, assuming only statistical uncertainties, $\epsilon_i =0$.

\subsection{Results}

\begin{figure}[h!]
	\centering
 	\subcaptionbox{Higgsino\label{fig:higgsino}}
	{\includegraphics[ width=0.49\textwidth]{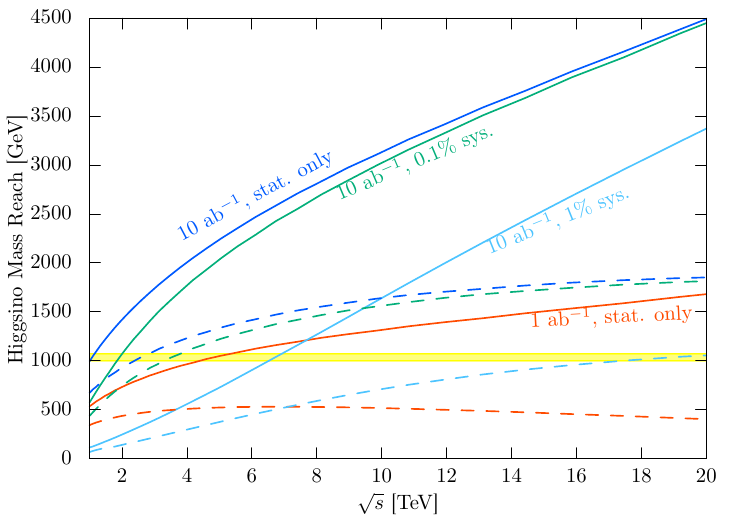}}
 	\subcaptionbox{ Wino  \label{fig:wino}}
	{\includegraphics[width=0.49\textwidth]{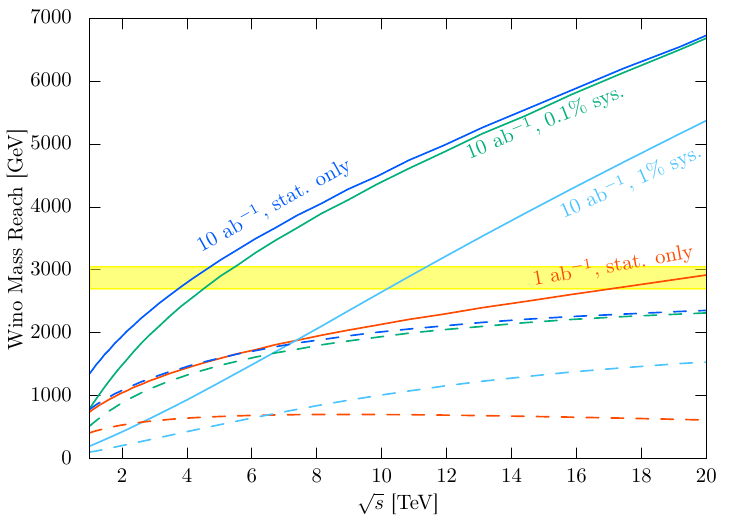}}
  	\subcaptionbox{ Minimal fermion DM  \label{fig:MDM}}
	{\includegraphics[width=0.49\textwidth]{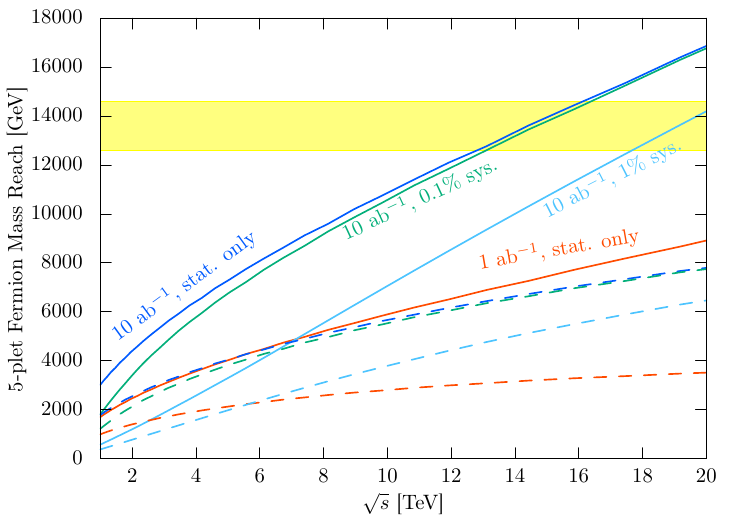}}
 	\subcaptionbox{ Minimal scalar DM  \label{fig:scMDM}}
	{\includegraphics[width=0.49\textwidth]{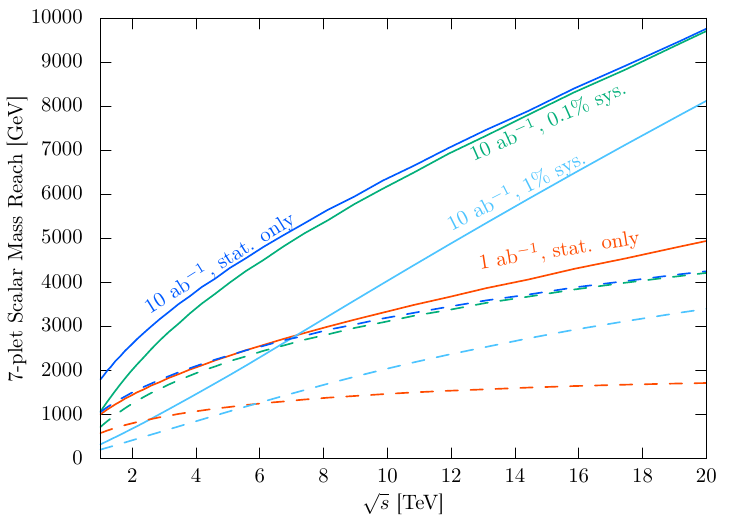}}
 \caption{
 The mass reach of EWIMP at 95\% CL.
 (a): Higgsino (Dirac fermion with $n=2,Y=\pm 1/2$).
 (b): Wino (Majorana fermion with $n=3,Y=0$).
 (c): Minimal fermion DM (Majorana fermion with $n=5,Y=0$)
 (d):  Minimal scalar DM (Real scalar with $n=7,Y=0$).
 The solid lines represent that the both initial anti-muons are right-handed ($P_{\mu^+}=+100$\%), while the dashed lines represent the unpolarized initial state ($P_{\mu^+}=0$\%).
 We have shown the plots with the systematic errors of $\epsilon_i = 0\%, 0.1\%, 1\%$ for the integrated luminocity of 10 ab$^{-1}$, while we have only plotted the case of $\epsilon_i=0\%$ for 1 ab$^{-1}$.
The yellow-shaded region denotes the range of EWIMP masses for which the thermal relic abundance of EWIMPs aligns with the observed DM abundance, $\Omega h^2 \simeq 0.12$~\cite{Hisano:2006nn,Cirelli:2005uq, *Cirelli:2007xd, *Cirelli:2009uv, Mitridate:2017izz,Bottaro:2021snn}.
Note that the predicted mass for the 7-plet real scalar DM is 54.2 $\pm$ 3.1 TeV~\cite{Bottaro:2021snn}, which is outside the scope of the figure.
}
 \label{fig:massreach}
\end{figure}

Here we consider four types of EWIMPs as benchmark points: Higgsino, Wino, 5-plet Minimal fermion DM (Majorana fermion with $n=5$ and $Y=0$), and 7-plet Minimal scalar DM (Real scalar with $n=7$ and $Y=0$).
For a given set of collider parameters, the collision energy $\sqrt{s}$, the integrated luminosity, and the systematic uncertainty $\epsilon_i$, we estimate $\chi^2$ given in Eq.\,\eqref{eq: chi2}.
We identify $\chi^2=3.8$ as the sensitivity of a 95\% confidence level.
Figure \ref{fig:massreach} shows the potential reach for detecting the mass of EWIMPs at a confidence level of 95\% through the precision measurement of the $\mu^+\mu^+\to\mu^+\mu^+$ scattering.
We show the cases  $(1~{\mathrm{ab}}^{-1}, \epsilon_i = 0)$ in red,  $(10~{\mathrm{ab}}^{-1}, \epsilon_i = 0)$ in blue,
$(10~{\mathrm{ab}}^{-1}, \epsilon_i = 0.001)$ in green, and $(10~{\mathrm{ab}}^{-1}, \epsilon_i = 0.01)$ in cyan. 
In each figure, the solid lines depict the scenario where both $\mu^+$ beams are entirely right-handed polarized, while the dashed lines represent the unpolarized case.
The yellow-shaded region represents the range of EWIMP masses for which the thermal relic abundance of EWIMPs aligns with the observed DM abundance, $\Omega h^2 \simeq 0.12$~\cite{Hisano:2006nn,Cirelli:2005uq, *Cirelli:2007xd, *Cirelli:2009uv}.

\begin{figure}[t]
    \centering
    \includegraphics[width=0.49\textwidth]{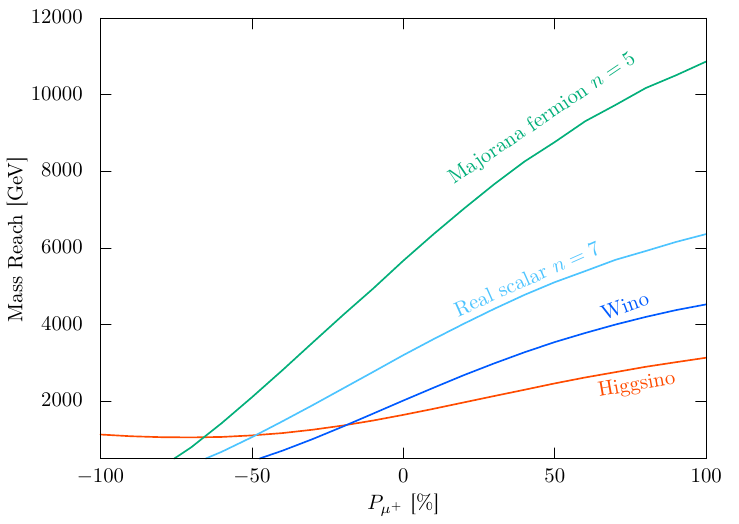}
    \caption{
The polarization dependence on the EWIMP mass reach. 
We assume a center-of-mass energy $\sqrt{s} = 10$ TeV, an integrated luminosity of $10\,\mathrm{ab}^{-1}$, and that systematic uncertainties are negligible with $\epsilon_i = 0$.
    }
    \label{fig:polarization}
\end{figure}

In Fig.\,\ref{fig:polarization}, we show the polarization dependence on the EWIMP mass reach, assuming a center-of-mass energy $\sqrt{s} = 10$ TeV, an integrated luminosity of $10\,\mathrm{ab}^{-1}$, and negligible systematic uncertainties with $\epsilon_i = 0$. 
The EWIMPs, except for the Higgsino, possess only $\mathrm{SU(2)}_L$ charges.
Consequently, the corrections diminish when the $\mu^+$ beams are left-handed polarized.

From these figures, we can see that the $\mu^+\mu^+$ collider with an energy of $\sqrt{s}\simeq 10-20$ TeV has the potential to investigate the thermal relic hypothesis of fermionic EWIMPs, provided that we maintain a very low systematic uncertainty of $O(0.1)$\% and utilize highly polarized beams.

%-----------------------------------------------------
\section{Summary and Discussion}
%-----------------------------------------------------

In this paper, we investigate the influence of EWIMPs on $\mu^+\mu^+\to\mu^+\mu^+$ scattering. We find that the EWIMP loop introduces a correction of $O(0.1-1)$\% to the SM process. Notably, this effect is particularly enhanced when the initial anti-muons are right-handed polarized. For the Higgsino and Wino EWIMPs, we identify that a beam energy of a few TeV and an integrated luminosity of $O(1)$ ab$^{-1}$ can be employed to search for mass values capable of explaining the correct thermal relic abundance.

Moreover, as depicted in Fig.\,\ref{fig:correction}, the scattering distribution of anti-muons depends on the quantum numbers of the EWIMP such as its spin, electroweak charge, and mass. 
Hence, the $\mu$TRISTAN experiment can play a role as an identifier of DM's properties by precisely measuring the scattering distribution.

It is intriguing to compare our results with other direct and indirect searches in future colliders.
As discussed in the introduction, experiments based on direct production of EWIMPs rely on detailed mass splitting of EWIMPs, and hence it is difficult to give a generic prospect. 
For simplicity, we consider the case of the pure Higgsino here. First, the disappearing track search by the High-Luminosity Large Hadron Collider (HL-LHC) has a sensitivity up to about 250 GeV for the pure Higgsino~\cite{ATLAS:2018jjf}.
The lepton-antilepton colliders such as $e^- e^+$ and $\mu^- \mu^+$ colliders can produce mono gauge bosons ($W$, $Z$, and $\gamma$) in addition to the EWIMP pair production, and the $\mu^- \mu^+$ collider is sensitive to the Higgsino of about 1.5 TeV using the mono-$W$ signal with $\sqrt{s} = 10$ TeV~\cite{Bottaro:2022one}.
The 10 TeV $\mu^- \mu^+$ ($\mu^+ \mu^+$) collider can also utilize the disappearing track and has a sensitivity up to about 1.5 TeV (1 TeV) for the pure Higgsino~\cite{Capdevilla:2021fmj,Kitano:2023pascos}.
The lepton-antilepton collider can also be used as an indirect search.
By utilizing the precision measurement of di-fermion production $\ell^- \ell^+ \to f \bar{f}$, where $f$ is the SM fermion, EWIMPs with masses around the beam energy can be detected~\cite{Harigaya:2015yaa}.

In this study, we have not considered beyond-one-loop corrections in the SM and BSM processes. By varying the renormalization scale from the EWIMP mass to $\sqrt{s}$, i.e., $\mu \in [m,\sqrt{s}]$, we estimated that corrections to the cross sections from EWIMPs beyond the one-loop correction are approximately $10$\%. This uncertainty introduces uncertainties of $\sim 10$\% to the mass reach.
For a more precise estimation, employing the resummation method, such as a muon parton distribution function, and subsequently matching it with the NLO hard process would be preferable.
However, such a precise estimation does not significantly affect the main assertion of our paper, which is the sensitivity estimate of indirect searchs for EWIMP at $\mu$TRISTAN.

Additionally, our approach artificially parametrizes the systematic uncertainty $\epsilon_i$ as a fixed value, whereas it should ideally be evaluated by considering uncertainties arising from both theoretical calculations and the experimental setup. We will address these considerations in future research.

The $\mu$TRISTAN experiment is still in the proposal stage, with detailed beam energies and detector designs yet to be finalized. Meanwhile, our present study offers a promising approach to investigate DM, a critical research objective. This work will significantly influence the design considerations for $\mu$TRISTAN.

\section*{Acknowledgements}

The work of R.O. was supported by the Forefront Physics and Mathematics Program to Drive Transformation (FoPM), a World-leading Innovative Graduate Study (WINGS) Program, the University of Tokyo, JSPS KAKENHI Grant Number JP24KJ0838, and the JSR Fellowship, the University of Tokyo.
The work of S.S. is supported by Grant-in-Aid for Scientific Research from the Ministry of Education, Culture, Sports, Science, and Technology (MEXT), Japan, 18K13535, 20H01895, 20H05860 and 21H00067.
This work is supported by World Premier International Research Center Initiative (WPI), MEXT, Japan.

{\it Note Added:} 
During the completion of this paper, a related study
\cite{Fukuda:2023yui} was submitted to arXiv which also discusses the indirect probe of the EWIMPs at the  $\mu^+\mu^+$ collider.

\bibliographystyle{apsrev4-1}
\bibliography{bibtex}

%merlin.mbs apsrev4-1.bst 2010-07-25 4.21a (PWD, AO, DPC) hacked
%Control: key (0)
%Control: author (72) initials jnrlst
%Control: editor formatted (1) identically to author
%Control: production of article title (-1) disabled
%Control: page (0) single
%Control: year (1) truncated
%Control: production of eprint (0) enabled
\begin{thebibliography}{41}%
\makeatletter
\providecommand \@ifxundefined [1]{%
 \@ifx{#1\undefined}
}%
\providecommand \@ifnum [1]{%
 \ifnum #1\expandafter \@firstoftwo
 \else \expandafter \@secondoftwo
 \fi
}%
\providecommand \@ifx [1]{%
 \ifx #1\expandafter \@firstoftwo
 \else \expandafter \@secondoftwo
 \fi
}%
\providecommand \natexlab [1]{#1}%
\providecommand \enquote  [1]{``#1''}%
\providecommand \bibnamefont  [1]{#1}%
\providecommand \bibfnamefont [1]{#1}%
\providecommand \citenamefont [1]{#1}%
\providecommand \href@noop [0]{\@secondoftwo}%
\providecommand \href [0]{\begingroup \@sanitize@url \@href}%
\providecommand \@href[1]{\@@startlink{#1}\@@href}%
\providecommand \@@href[1]{\endgroup#1\@@endlink}%
\providecommand \@sanitize@url [0]{\catcode `\\12\catcode `\$12\catcode `\&12\catcode `\#12\catcode `\^12\catcode `\_12\catcode `\%12\relax}%
\providecommand \@@startlink[1]{}%
\providecommand \@@endlink[0]{}%
\providecommand \url  [0]{\begingroup\@sanitize@url \@url }%
\providecommand \@url [1]{\endgroup\@href {#1}{\urlprefix }}%
\providecommand \urlprefix  [0]{URL }%
\providecommand \Eprint [0]{\href }%
\providecommand \doibase [0]{http://dx.doi.org/}%
\providecommand \selectlanguage [0]{\@gobble}%
\providecommand \bibinfo  [0]{\@secondoftwo}%
\providecommand \bibfield  [0]{\@secondoftwo}%
\providecommand \translation [1]{[#1]}%
\providecommand \BibitemOpen [0]{}%
\providecommand \bibitemStop [0]{}%
\providecommand \bibitemNoStop [0]{.\EOS\space}%
\providecommand \EOS [0]{\spacefactor3000\relax}%
\providecommand \BibitemShut  [1]{\csname bibitem#1\endcsname}%
\let\auto@bib@innerbib\@empty
%</preamble>
\bibitem [{\citenamefont {Cirelli}\ \emph {et~al.}(2006)\citenamefont {Cirelli}, \citenamefont {Fornengo},\ and\ \citenamefont {Strumia}}]{Cirelli:2005uq}%
  \BibitemOpen
  \bibfield  {author} {\bibinfo {author} {\bibfnamefont {M.}~\bibnamefont {Cirelli}}, \bibinfo {author} {\bibfnamefont {N.}~\bibnamefont {Fornengo}}, \ and\ \bibinfo {author} {\bibfnamefont {A.}~\bibnamefont {Strumia}},\ }\href {\doibase 10.1016/j.nuclphysb.2006.07.012} {\bibfield  {journal} {\bibinfo  {journal} {Nucl. Phys. B}\ }\textbf {\bibinfo {volume} {753}},\ \bibinfo {pages} {178} (\bibinfo {year} {2006})},\ \Eprint {http://arxiv.org/abs/hep-ph/0512090} {arXiv:hep-ph/0512090} \BibitemShut {NoStop}%
\bibitem [{\citenamefont {Cirelli}\ \emph {et~al.}(2007)\citenamefont {Cirelli}, \citenamefont {Strumia},\ and\ \citenamefont {Tamburini}}]{Cirelli:2007xd}%
  \BibitemOpen
  \bibfield  {author} {\bibinfo {author} {\bibfnamefont {M.}~\bibnamefont {Cirelli}}, \bibinfo {author} {\bibfnamefont {A.}~\bibnamefont {Strumia}}, \ and\ \bibinfo {author} {\bibfnamefont {M.}~\bibnamefont {Tamburini}},\ }\href {\doibase 10.1016/j.nuclphysb.2007.07.023} {\bibfield  {journal} {\bibinfo  {journal} {Nucl. Phys. B}\ }\textbf {\bibinfo {volume} {787}},\ \bibinfo {pages} {152} (\bibinfo {year} {2007})},\ \Eprint {http://arxiv.org/abs/0706.4071} {arXiv:0706.4071 [hep-ph]} \BibitemShut {NoStop}%
\bibitem [{\citenamefont {Cirelli}\ and\ \citenamefont {Strumia}(2009)}]{Cirelli:2009uv}%
  \BibitemOpen
  \bibfield  {author} {\bibinfo {author} {\bibfnamefont {M.}~\bibnamefont {Cirelli}}\ and\ \bibinfo {author} {\bibfnamefont {A.}~\bibnamefont {Strumia}},\ }\href {\doibase 10.1088/1367-2630/11/10/105005} {\bibfield  {journal} {\bibinfo  {journal} {New J. Phys.}\ }\textbf {\bibinfo {volume} {11}},\ \bibinfo {pages} {105005} (\bibinfo {year} {2009})},\ \Eprint {http://arxiv.org/abs/0903.3381} {arXiv:0903.3381 [hep-ph]} \BibitemShut {NoStop}%
\bibitem [{\citenamefont {Chen}\ \emph {et~al.}(1997)\citenamefont {Chen}, \citenamefont {Drees},\ and\ \citenamefont {Gunion}}]{Chen:1996ap}%
  \BibitemOpen
  \bibfield  {author} {\bibinfo {author} {\bibfnamefont {C.~H.}\ \bibnamefont {Chen}}, \bibinfo {author} {\bibfnamefont {M.}~\bibnamefont {Drees}}, \ and\ \bibinfo {author} {\bibfnamefont {J.~F.}\ \bibnamefont {Gunion}},\ }\href {\doibase 10.1103/PhysRevD.60.039901} {\bibfield  {journal} {\bibinfo  {journal} {Phys. Rev. D}\ }\textbf {\bibinfo {volume} {55}},\ \bibinfo {pages} {330} (\bibinfo {year} {1997})},\ \bibinfo {note} {[Erratum: Phys.Rev.D 60, 039901 (1999)]},\ \Eprint {http://arxiv.org/abs/hep-ph/9607421} {arXiv:hep-ph/9607421} \BibitemShut {NoStop}%
\bibitem [{\citenamefont {Thomas}\ and\ \citenamefont {Wells}(1998)}]{Thomas:1998wy}%
  \BibitemOpen
  \bibfield  {author} {\bibinfo {author} {\bibfnamefont {S.~D.}\ \bibnamefont {Thomas}}\ and\ \bibinfo {author} {\bibfnamefont {J.~D.}\ \bibnamefont {Wells}},\ }\href {\doibase 10.1103/PhysRevLett.81.34} {\bibfield  {journal} {\bibinfo  {journal} {Phys. Rev. Lett.}\ }\textbf {\bibinfo {volume} {81}},\ \bibinfo {pages} {34} (\bibinfo {year} {1998})},\ \Eprint {http://arxiv.org/abs/hep-ph/9804359} {arXiv:hep-ph/9804359} \BibitemShut {NoStop}%
\bibitem [{\citenamefont {Nagata}\ and\ \citenamefont {Shirai}(2015)}]{Nagata:2014wma}%
  \BibitemOpen
  \bibfield  {author} {\bibinfo {author} {\bibfnamefont {N.}~\bibnamefont {Nagata}}\ and\ \bibinfo {author} {\bibfnamefont {S.}~\bibnamefont {Shirai}},\ }\href {\doibase 10.1007/JHEP01(2015)029} {\bibfield  {journal} {\bibinfo  {journal} {JHEP}\ }\textbf {\bibinfo {volume} {01}},\ \bibinfo {pages} {029} (\bibinfo {year} {2015})},\ \Eprint {http://arxiv.org/abs/1410.4549} {arXiv:1410.4549 [hep-ph]} \BibitemShut {NoStop}%
\bibitem [{\citenamefont {Ibe}\ \emph {et~al.}(2023)\citenamefont {Ibe}, \citenamefont {Mishima}, \citenamefont {Nakayama},\ and\ \citenamefont {Shirai}}]{Ibe:2022lkl}%
  \BibitemOpen
  \bibfield  {author} {\bibinfo {author} {\bibfnamefont {M.}~\bibnamefont {Ibe}}, \bibinfo {author} {\bibfnamefont {M.}~\bibnamefont {Mishima}}, \bibinfo {author} {\bibfnamefont {Y.}~\bibnamefont {Nakayama}}, \ and\ \bibinfo {author} {\bibfnamefont {S.}~\bibnamefont {Shirai}},\ }\href {\doibase 10.1007/JHEP01(2023)017} {\bibfield  {journal} {\bibinfo  {journal} {JHEP}\ }\textbf {\bibinfo {volume} {01}},\ \bibinfo {pages} {017} (\bibinfo {year} {2023})},\ \Eprint {http://arxiv.org/abs/2210.16035} {arXiv:2210.16035 [hep-ph]} \BibitemShut {NoStop}%
\bibitem [{\citenamefont {Ibe}\ \emph {et~al.}(2007)\citenamefont {Ibe}, \citenamefont {Moroi},\ and\ \citenamefont {Yanagida}}]{Ibe:2006de}%
  \BibitemOpen
  \bibfield  {author} {\bibinfo {author} {\bibfnamefont {M.}~\bibnamefont {Ibe}}, \bibinfo {author} {\bibfnamefont {T.}~\bibnamefont {Moroi}}, \ and\ \bibinfo {author} {\bibfnamefont {T.~T.}\ \bibnamefont {Yanagida}},\ }\href {\doibase 10.1016/j.physletb.2006.11.061} {\bibfield  {journal} {\bibinfo  {journal} {Phys. Lett. B}\ }\textbf {\bibinfo {volume} {644}},\ \bibinfo {pages} {355} (\bibinfo {year} {2007})},\ \Eprint {http://arxiv.org/abs/hep-ph/0610277} {arXiv:hep-ph/0610277} \BibitemShut {NoStop}%
\bibitem [{\citenamefont {Buckley}\ \emph {et~al.}(2011)\citenamefont {Buckley}, \citenamefont {Randall},\ and\ \citenamefont {Shuve}}]{Buckley:2009kv}%
  \BibitemOpen
  \bibfield  {author} {\bibinfo {author} {\bibfnamefont {M.~R.}\ \bibnamefont {Buckley}}, \bibinfo {author} {\bibfnamefont {L.}~\bibnamefont {Randall}}, \ and\ \bibinfo {author} {\bibfnamefont {B.}~\bibnamefont {Shuve}},\ }\href {\doibase 10.1007/JHEP05(2011)097} {\bibfield  {journal} {\bibinfo  {journal} {JHEP}\ }\textbf {\bibinfo {volume} {05}},\ \bibinfo {pages} {097} (\bibinfo {year} {2011})},\ \Eprint {http://arxiv.org/abs/0909.4549} {arXiv:0909.4549 [hep-ph]} \BibitemShut {NoStop}%
\bibitem [{\citenamefont {Asai}\ \emph {et~al.}(2007)\citenamefont {Asai}, \citenamefont {Moroi}, \citenamefont {Nishihara},\ and\ \citenamefont {Yanagida}}]{Asai:2007sw}%
  \BibitemOpen
  \bibfield  {author} {\bibinfo {author} {\bibfnamefont {S.}~\bibnamefont {Asai}}, \bibinfo {author} {\bibfnamefont {T.}~\bibnamefont {Moroi}}, \bibinfo {author} {\bibfnamefont {K.}~\bibnamefont {Nishihara}}, \ and\ \bibinfo {author} {\bibfnamefont {T.~T.}\ \bibnamefont {Yanagida}},\ }\href {\doibase 10.1016/j.physletb.2007.06.080} {\bibfield  {journal} {\bibinfo  {journal} {Phys. Lett. B}\ }\textbf {\bibinfo {volume} {653}},\ \bibinfo {pages} {81} (\bibinfo {year} {2007})},\ \Eprint {http://arxiv.org/abs/0705.3086} {arXiv:0705.3086 [hep-ph]} \BibitemShut {NoStop}%
\bibitem [{\citenamefont {Asai}\ \emph {et~al.}(2008)\citenamefont {Asai}, \citenamefont {Moroi},\ and\ \citenamefont {Yanagida}}]{Asai:2008sk}%
  \BibitemOpen
  \bibfield  {author} {\bibinfo {author} {\bibfnamefont {S.}~\bibnamefont {Asai}}, \bibinfo {author} {\bibfnamefont {T.}~\bibnamefont {Moroi}}, \ and\ \bibinfo {author} {\bibfnamefont {T.~T.}\ \bibnamefont {Yanagida}},\ }\href {\doibase 10.1016/j.physletb.2008.05.019} {\bibfield  {journal} {\bibinfo  {journal} {Phys. Lett. B}\ }\textbf {\bibinfo {volume} {664}},\ \bibinfo {pages} {185} (\bibinfo {year} {2008})},\ \Eprint {http://arxiv.org/abs/0802.3725} {arXiv:0802.3725 [hep-ph]} \BibitemShut {NoStop}%
\bibitem [{\citenamefont {Asai}\ \emph {et~al.}(2009)\citenamefont {Asai}, \citenamefont {Azuma}, \citenamefont {Jinnouchi}, \citenamefont {Moroi}, \citenamefont {Shirai},\ and\ \citenamefont {Yanagida}}]{Asai:2008im}%
  \BibitemOpen
  \bibfield  {author} {\bibinfo {author} {\bibfnamefont {S.}~\bibnamefont {Asai}}, \bibinfo {author} {\bibfnamefont {Y.}~\bibnamefont {Azuma}}, \bibinfo {author} {\bibfnamefont {O.}~\bibnamefont {Jinnouchi}}, \bibinfo {author} {\bibfnamefont {T.}~\bibnamefont {Moroi}}, \bibinfo {author} {\bibfnamefont {S.}~\bibnamefont {Shirai}}, \ and\ \bibinfo {author} {\bibfnamefont {T.~T.}\ \bibnamefont {Yanagida}},\ }\href {\doibase 10.1016/j.physletb.2009.01.045} {\bibfield  {journal} {\bibinfo  {journal} {Phys. Lett. B}\ }\textbf {\bibinfo {volume} {672}},\ \bibinfo {pages} {339} (\bibinfo {year} {2009})},\ \Eprint {http://arxiv.org/abs/0807.4987} {arXiv:0807.4987 [hep-ph]} \BibitemShut {NoStop}%
\bibitem [{\citenamefont {Mahbubani}\ \emph {et~al.}(2017)\citenamefont {Mahbubani}, \citenamefont {Schwaller},\ and\ \citenamefont {Zurita}}]{Mahbubani:2017gjh}%
  \BibitemOpen
  \bibfield  {author} {\bibinfo {author} {\bibfnamefont {R.}~\bibnamefont {Mahbubani}}, \bibinfo {author} {\bibfnamefont {P.}~\bibnamefont {Schwaller}}, \ and\ \bibinfo {author} {\bibfnamefont {J.}~\bibnamefont {Zurita}},\ }\href {\doibase 10.1007/JHEP06(2017)119} {\bibfield  {journal} {\bibinfo  {journal} {JHEP}\ }\textbf {\bibinfo {volume} {06}},\ \bibinfo {pages} {119} (\bibinfo {year} {2017})},\ \bibinfo {note} {[Erratum: JHEP 10, 061 (2017)]},\ \Eprint {http://arxiv.org/abs/1703.05327} {arXiv:1703.05327 [hep-ph]} \BibitemShut {NoStop}%
\bibitem [{\citenamefont {Fukuda}\ \emph {et~al.}(2018)\citenamefont {Fukuda}, \citenamefont {Nagata}, \citenamefont {Otono},\ and\ \citenamefont {Shirai}}]{Fukuda:2017jmk}%
  \BibitemOpen
  \bibfield  {author} {\bibinfo {author} {\bibfnamefont {H.}~\bibnamefont {Fukuda}}, \bibinfo {author} {\bibfnamefont {N.}~\bibnamefont {Nagata}}, \bibinfo {author} {\bibfnamefont {H.}~\bibnamefont {Otono}}, \ and\ \bibinfo {author} {\bibfnamefont {S.}~\bibnamefont {Shirai}},\ }\href {\doibase 10.1016/j.physletb.2018.03.088} {\bibfield  {journal} {\bibinfo  {journal} {Phys. Lett. B}\ }\textbf {\bibinfo {volume} {781}},\ \bibinfo {pages} {306} (\bibinfo {year} {2018})},\ \Eprint {http://arxiv.org/abs/1703.09675} {arXiv:1703.09675 [hep-ph]} \BibitemShut {NoStop}%
\bibitem [{\citenamefont {Fukuda}\ \emph {et~al.}(2020)\citenamefont {Fukuda}, \citenamefont {Nagata}, \citenamefont {Oide}, \citenamefont {Otono},\ and\ \citenamefont {Shirai}}]{Fukuda:2019kbp}%
  \BibitemOpen
  \bibfield  {author} {\bibinfo {author} {\bibfnamefont {H.}~\bibnamefont {Fukuda}}, \bibinfo {author} {\bibfnamefont {N.}~\bibnamefont {Nagata}}, \bibinfo {author} {\bibfnamefont {H.}~\bibnamefont {Oide}}, \bibinfo {author} {\bibfnamefont {H.}~\bibnamefont {Otono}}, \ and\ \bibinfo {author} {\bibfnamefont {S.}~\bibnamefont {Shirai}},\ }\href {\doibase 10.1103/PhysRevLett.124.101801} {\bibfield  {journal} {\bibinfo  {journal} {Phys. Rev. Lett.}\ }\textbf {\bibinfo {volume} {124}},\ \bibinfo {pages} {101801} (\bibinfo {year} {2020})},\ \Eprint {http://arxiv.org/abs/1910.08065} {arXiv:1910.08065 [hep-ph]} \BibitemShut {NoStop}%
\bibitem [{\citenamefont {Ibe}\ \emph {et~al.}(2024)\citenamefont {Ibe}, \citenamefont {Nakayama},\ and\ \citenamefont {Shirai}}]{Ibe:2023dcu}%
  \BibitemOpen
  \bibfield  {author} {\bibinfo {author} {\bibfnamefont {M.}~\bibnamefont {Ibe}}, \bibinfo {author} {\bibfnamefont {Y.}~\bibnamefont {Nakayama}}, \ and\ \bibinfo {author} {\bibfnamefont {S.}~\bibnamefont {Shirai}},\ }\href {\doibase 10.1007/JHEP03(2024)012} {\bibfield  {journal} {\bibinfo  {journal} {JHEP}\ }\textbf {\bibinfo {volume} {03}},\ \bibinfo {pages} {012} (\bibinfo {year} {2024})},\ \Eprint {http://arxiv.org/abs/2312.08087} {arXiv:2312.08087 [hep-ph]} \BibitemShut {NoStop}%
\bibitem [{\citenamefont {Aad}\ \emph {et~al.}(2022)\citenamefont {Aad} \emph {et~al.}}]{ATLAS:2022rme}%
  \BibitemOpen
  \bibfield  {author} {\bibinfo {author} {\bibfnamefont {G.}~\bibnamefont {Aad}} \emph {et~al.} (\bibinfo {collaboration} {ATLAS}),\ }\href {\doibase 10.1140/epjc/s10052-022-10489-5} {\bibfield  {journal} {\bibinfo  {journal} {Eur. Phys. J. C}\ }\textbf {\bibinfo {volume} {82}},\ \bibinfo {pages} {606} (\bibinfo {year} {2022})},\ \Eprint {http://arxiv.org/abs/2201.02472} {arXiv:2201.02472 [hep-ex]} \BibitemShut {NoStop}%
\bibitem [{\citenamefont {Harigaya}\ \emph {et~al.}(2015)\citenamefont {Harigaya}, \citenamefont {Ichikawa}, \citenamefont {Kundu}, \citenamefont {Matsumoto},\ and\ \citenamefont {Shirai}}]{Harigaya:2015yaa}%
  \BibitemOpen
  \bibfield  {author} {\bibinfo {author} {\bibfnamefont {K.}~\bibnamefont {Harigaya}}, \bibinfo {author} {\bibfnamefont {K.}~\bibnamefont {Ichikawa}}, \bibinfo {author} {\bibfnamefont {A.}~\bibnamefont {Kundu}}, \bibinfo {author} {\bibfnamefont {S.}~\bibnamefont {Matsumoto}}, \ and\ \bibinfo {author} {\bibfnamefont {S.}~\bibnamefont {Shirai}},\ }\href {\doibase 10.1007/JHEP09(2015)105} {\bibfield  {journal} {\bibinfo  {journal} {JHEP}\ }\textbf {\bibinfo {volume} {09}},\ \bibinfo {pages} {105} (\bibinfo {year} {2015})},\ \Eprint {http://arxiv.org/abs/1504.03402} {arXiv:1504.03402 [hep-ph]} \BibitemShut {NoStop}%
\bibitem [{\citenamefont {Franceschini}\ and\ \citenamefont {Zhao}(2023)}]{Franceschini:2022sxc}%
  \BibitemOpen
  \bibfield  {author} {\bibinfo {author} {\bibfnamefont {R.}~\bibnamefont {Franceschini}}\ and\ \bibinfo {author} {\bibfnamefont {X.}~\bibnamefont {Zhao}},\ }\href {\doibase 10.1140/epjc/s10052-023-11724-3} {\bibfield  {journal} {\bibinfo  {journal} {Eur. Phys. J. C}\ }\textbf {\bibinfo {volume} {83}},\ \bibinfo {pages} {552} (\bibinfo {year} {2023})},\ \Eprint {http://arxiv.org/abs/2212.11900} {arXiv:2212.11900 [hep-ph]} \BibitemShut {NoStop}%
\bibitem [{\citenamefont {Matsumoto}\ \emph {et~al.}(2018)\citenamefont {Matsumoto}, \citenamefont {Shirai},\ and\ \citenamefont {Takeuchi}}]{Matsumoto:2017vfu}%
  \BibitemOpen
  \bibfield  {author} {\bibinfo {author} {\bibfnamefont {S.}~\bibnamefont {Matsumoto}}, \bibinfo {author} {\bibfnamefont {S.}~\bibnamefont {Shirai}}, \ and\ \bibinfo {author} {\bibfnamefont {M.}~\bibnamefont {Takeuchi}},\ }\href {\doibase 10.1007/JHEP06(2018)049} {\bibfield  {journal} {\bibinfo  {journal} {JHEP}\ }\textbf {\bibinfo {volume} {06}},\ \bibinfo {pages} {049} (\bibinfo {year} {2018})},\ \Eprint {http://arxiv.org/abs/1711.05449} {arXiv:1711.05449 [hep-ph]} \BibitemShut {NoStop}%
\bibitem [{\citenamefont {Chigusa}\ \emph {et~al.}(2019)\citenamefont {Chigusa}, \citenamefont {Ema},\ and\ \citenamefont {Moroi}}]{Chigusa:2018vxz}%
  \BibitemOpen
  \bibfield  {author} {\bibinfo {author} {\bibfnamefont {S.}~\bibnamefont {Chigusa}}, \bibinfo {author} {\bibfnamefont {Y.}~\bibnamefont {Ema}}, \ and\ \bibinfo {author} {\bibfnamefont {T.}~\bibnamefont {Moroi}},\ }\href {\doibase 10.1016/j.physletb.2018.12.011} {\bibfield  {journal} {\bibinfo  {journal} {Phys. Lett. B}\ }\textbf {\bibinfo {volume} {789}},\ \bibinfo {pages} {106} (\bibinfo {year} {2019})},\ \Eprint {http://arxiv.org/abs/1810.07349} {arXiv:1810.07349 [hep-ph]} \BibitemShut {NoStop}%
\bibitem [{\citenamefont {Di~Luzio}\ \emph {et~al.}(2019)\citenamefont {Di~Luzio}, \citenamefont {Gr\"ober},\ and\ \citenamefont {Panico}}]{DiLuzio:2018jwd}%
  \BibitemOpen
  \bibfield  {author} {\bibinfo {author} {\bibfnamefont {L.}~\bibnamefont {Di~Luzio}}, \bibinfo {author} {\bibfnamefont {R.}~\bibnamefont {Gr\"ober}}, \ and\ \bibinfo {author} {\bibfnamefont {G.}~\bibnamefont {Panico}},\ }\href {\doibase 10.1007/JHEP01(2019)011} {\bibfield  {journal} {\bibinfo  {journal} {JHEP}\ }\textbf {\bibinfo {volume} {01}},\ \bibinfo {pages} {011} (\bibinfo {year} {2019})},\ \Eprint {http://arxiv.org/abs/1810.10993} {arXiv:1810.10993 [hep-ph]} \BibitemShut {NoStop}%
\bibitem [{\citenamefont {Matsumoto}\ \emph {et~al.}(2019)\citenamefont {Matsumoto}, \citenamefont {Shirai},\ and\ \citenamefont {Takeuchi}}]{Matsumoto:2018ioi}%
  \BibitemOpen
  \bibfield  {author} {\bibinfo {author} {\bibfnamefont {S.}~\bibnamefont {Matsumoto}}, \bibinfo {author} {\bibfnamefont {S.}~\bibnamefont {Shirai}}, \ and\ \bibinfo {author} {\bibfnamefont {M.}~\bibnamefont {Takeuchi}},\ }\href {\doibase 10.1007/JHEP03(2019)076} {\bibfield  {journal} {\bibinfo  {journal} {JHEP}\ }\textbf {\bibinfo {volume} {03}},\ \bibinfo {pages} {076} (\bibinfo {year} {2019})},\ \Eprint {http://arxiv.org/abs/1810.12234} {arXiv:1810.12234 [hep-ph]} \BibitemShut {NoStop}%
\bibitem [{\citenamefont {Katayose}\ \emph {et~al.}(2021)\citenamefont {Katayose}, \citenamefont {Matsumoto},\ and\ \citenamefont {Shirai}}]{Katayose:2020one}%
  \BibitemOpen
  \bibfield  {author} {\bibinfo {author} {\bibfnamefont {T.}~\bibnamefont {Katayose}}, \bibinfo {author} {\bibfnamefont {S.}~\bibnamefont {Matsumoto}}, \ and\ \bibinfo {author} {\bibfnamefont {S.}~\bibnamefont {Shirai}},\ }\href {\doibase 10.1103/PhysRevD.103.095017} {\bibfield  {journal} {\bibinfo  {journal} {Phys. Rev. D}\ }\textbf {\bibinfo {volume} {103}},\ \bibinfo {pages} {095017} (\bibinfo {year} {2021})},\ \Eprint {http://arxiv.org/abs/2011.14784} {arXiv:2011.14784 [hep-ph]} \BibitemShut {NoStop}%
\bibitem [{\citenamefont {Hamada}\ \emph {et~al.}(2022)\citenamefont {Hamada}, \citenamefont {Kitano}, \citenamefont {Matsudo}, \citenamefont {Takaura},\ and\ \citenamefont {Yoshida}}]{Hamada:2022mua}%
  \BibitemOpen
  \bibfield  {author} {\bibinfo {author} {\bibfnamefont {Y.}~\bibnamefont {Hamada}}, \bibinfo {author} {\bibfnamefont {R.}~\bibnamefont {Kitano}}, \bibinfo {author} {\bibfnamefont {R.}~\bibnamefont {Matsudo}}, \bibinfo {author} {\bibfnamefont {H.}~\bibnamefont {Takaura}}, \ and\ \bibinfo {author} {\bibfnamefont {M.}~\bibnamefont {Yoshida}},\ }\href {\doibase 10.1093/ptep/ptac059} {\bibfield  {journal} {\bibinfo  {journal} {PTEP}\ }\textbf {\bibinfo {volume} {2022}},\ \bibinfo {pages} {053B02} (\bibinfo {year} {2022})},\ \Eprint {http://arxiv.org/abs/2201.06664} {arXiv:2201.06664 [hep-ph]} \BibitemShut {NoStop}%
\bibitem [{\citenamefont {Abe}\ \emph {et~al.}(2019)\citenamefont {Abe} \emph {et~al.}}]{Abe:2019thb}%
  \BibitemOpen
  \bibfield  {author} {\bibinfo {author} {\bibfnamefont {M.}~\bibnamefont {Abe}} \emph {et~al.},\ }\href {\doibase 10.1093/ptep/ptz030} {\bibfield  {journal} {\bibinfo  {journal} {PTEP}\ }\textbf {\bibinfo {volume} {2019}},\ \bibinfo {pages} {053C02} (\bibinfo {year} {2019})},\ \Eprint {http://arxiv.org/abs/1901.03047} {arXiv:1901.03047 [physics.ins-det]} \BibitemShut {NoStop}%
\bibitem [{\citenamefont {Kitano}(2023)}]{Kitano:2023pascos}%
  \BibitemOpen
  \bibfield  {author} {\bibinfo {author} {\bibfnamefont {R.}~\bibnamefont {Kitano}},\ }\href {https://indico.cern.ch/event/1188534/contributions/5438451/attachments/2675487/4639577/PASCOS_062923_kitano.pdf} {\enquote {\bibinfo {title} {{$\mu$TRISTAN}},}\ } (\bibinfo {year} {2023}),\ \bibinfo {note} {{PASCOS 2023}}\BibitemShut {NoStop}%
\bibitem [{\citenamefont {Hamada}\ \emph {et~al.}(2023)\citenamefont {Hamada}, \citenamefont {Kitano}, \citenamefont {Matsudo},\ and\ \citenamefont {Takaura}}]{Hamada:2022uyn}%
  \BibitemOpen
  \bibfield  {author} {\bibinfo {author} {\bibfnamefont {Y.}~\bibnamefont {Hamada}}, \bibinfo {author} {\bibfnamefont {R.}~\bibnamefont {Kitano}}, \bibinfo {author} {\bibfnamefont {R.}~\bibnamefont {Matsudo}}, \ and\ \bibinfo {author} {\bibfnamefont {H.}~\bibnamefont {Takaura}},\ }\href {\doibase 10.1093/ptep/ptac174} {\bibfield  {journal} {\bibinfo  {journal} {PTEP}\ }\textbf {\bibinfo {volume} {2023}},\ \bibinfo {pages} {013B07} (\bibinfo {year} {2023})},\ \Eprint {http://arxiv.org/abs/2210.11083} {arXiv:2210.11083 [hep-ph]} \BibitemShut {NoStop}%
\bibitem [{\citenamefont {Lorca}\ and\ \citenamefont {Riemann}(2006)}]{Lorca:2004fg}%
  \BibitemOpen
  \bibfield  {author} {\bibinfo {author} {\bibfnamefont {A.}~\bibnamefont {Lorca}}\ and\ \bibinfo {author} {\bibfnamefont {T.}~\bibnamefont {Riemann}},\ }\href {\doibase 10.1016/j.cpc.2005.09.003} {\bibfield  {journal} {\bibinfo  {journal} {Comput. Phys. Commun.}\ }\textbf {\bibinfo {volume} {174}},\ \bibinfo {pages} {71} (\bibinfo {year} {2006})},\ \Eprint {http://arxiv.org/abs/hep-ph/0412047} {arXiv:hep-ph/0412047} \BibitemShut {NoStop}%
\bibitem [{\citenamefont {Nogueira}(1993)}]{Nogueira:1991ex}%
  \BibitemOpen
  \bibfield  {author} {\bibinfo {author} {\bibfnamefont {P.}~\bibnamefont {Nogueira}},\ }\href {\doibase 10.1006/jcph.1993.1074} {\bibfield  {journal} {\bibinfo  {journal} {J. Comput. Phys.}\ }\textbf {\bibinfo {volume} {105}},\ \bibinfo {pages} {279} (\bibinfo {year} {1993})}\BibitemShut {NoStop}%
\bibitem [{\citenamefont {Tentyukov}\ and\ \citenamefont {Fleischer}(2000)}]{Tentyukov:1999is}%
  \BibitemOpen
  \bibfield  {author} {\bibinfo {author} {\bibfnamefont {M.}~\bibnamefont {Tentyukov}}\ and\ \bibinfo {author} {\bibfnamefont {J.}~\bibnamefont {Fleischer}},\ }\href {\doibase 10.1016/S0010-4655(00)00147-8} {\bibfield  {journal} {\bibinfo  {journal} {Comput. Phys. Commun.}\ }\textbf {\bibinfo {volume} {132}},\ \bibinfo {pages} {124} (\bibinfo {year} {2000})},\ \Eprint {http://arxiv.org/abs/hep-ph/9904258} {arXiv:hep-ph/9904258} \BibitemShut {NoStop}%
\bibitem [{\citenamefont {Vermaseren}(2000)}]{Vermaseren:2000nd}%
  \BibitemOpen
  \bibfield  {author} {\bibinfo {author} {\bibfnamefont {J.~A.~M.}\ \bibnamefont {Vermaseren}},\ }\href@noop {} {\  (\bibinfo {year} {2000})},\ \Eprint {http://arxiv.org/abs/math-ph/0010025} {arXiv:math-ph/0010025} \BibitemShut {NoStop}%
\bibitem [{\citenamefont {Hahn}\ and\ \citenamefont {Perez-Victoria}(1999)}]{Hahn:1998yk}%
  \BibitemOpen
  \bibfield  {author} {\bibinfo {author} {\bibfnamefont {T.}~\bibnamefont {Hahn}}\ and\ \bibinfo {author} {\bibfnamefont {M.}~\bibnamefont {Perez-Victoria}},\ }\href {\doibase 10.1016/S0010-4655(98)00173-8} {\bibfield  {journal} {\bibinfo  {journal} {Comput. Phys. Commun.}\ }\textbf {\bibinfo {volume} {118}},\ \bibinfo {pages} {153} (\bibinfo {year} {1999})},\ \Eprint {http://arxiv.org/abs/hep-ph/9807565} {arXiv:hep-ph/9807565} \BibitemShut {NoStop}%
\bibitem [{\citenamefont {van Oldenborgh}(1991)}]{vanOldenborgh:1990yc}%
  \BibitemOpen
  \bibfield  {author} {\bibinfo {author} {\bibfnamefont {G.~J.}\ \bibnamefont {van Oldenborgh}},\ }\href {\doibase 10.1016/0010-4655(91)90002-3} {\bibfield  {journal} {\bibinfo  {journal} {Comput. Phys. Commun.}\ }\textbf {\bibinfo {volume} {66}},\ \bibinfo {pages} {1} (\bibinfo {year} {1991})}\BibitemShut {NoStop}%
\bibitem [{\citenamefont {Hisano}\ \emph {et~al.}(2007)\citenamefont {Hisano}, \citenamefont {Matsumoto}, \citenamefont {Nagai}, \citenamefont {Saito},\ and\ \citenamefont {Senami}}]{Hisano:2006nn}%
  \BibitemOpen
  \bibfield  {author} {\bibinfo {author} {\bibfnamefont {J.}~\bibnamefont {Hisano}}, \bibinfo {author} {\bibfnamefont {S.}~\bibnamefont {Matsumoto}}, \bibinfo {author} {\bibfnamefont {M.}~\bibnamefont {Nagai}}, \bibinfo {author} {\bibfnamefont {O.}~\bibnamefont {Saito}}, \ and\ \bibinfo {author} {\bibfnamefont {M.}~\bibnamefont {Senami}},\ }\href {\doibase 10.1016/j.physletb.2007.01.012} {\bibfield  {journal} {\bibinfo  {journal} {Phys. Lett. B}\ }\textbf {\bibinfo {volume} {646}},\ \bibinfo {pages} {34} (\bibinfo {year} {2007})},\ \Eprint {http://arxiv.org/abs/hep-ph/0610249} {arXiv:hep-ph/0610249} \BibitemShut {NoStop}%
\bibitem [{\citenamefont {Mitridate}\ \emph {et~al.}(2017)\citenamefont {Mitridate}, \citenamefont {Redi}, \citenamefont {Smirnov},\ and\ \citenamefont {Strumia}}]{Mitridate:2017izz}%
  \BibitemOpen
  \bibfield  {author} {\bibinfo {author} {\bibfnamefont {A.}~\bibnamefont {Mitridate}}, \bibinfo {author} {\bibfnamefont {M.}~\bibnamefont {Redi}}, \bibinfo {author} {\bibfnamefont {J.}~\bibnamefont {Smirnov}}, \ and\ \bibinfo {author} {\bibfnamefont {A.}~\bibnamefont {Strumia}},\ }\href {\doibase 10.1088/1475-7516/2017/05/006} {\bibfield  {journal} {\bibinfo  {journal} {JCAP}\ }\textbf {\bibinfo {volume} {05}},\ \bibinfo {pages} {006} (\bibinfo {year} {2017})},\ \Eprint {http://arxiv.org/abs/1702.01141} {arXiv:1702.01141 [hep-ph]} \BibitemShut {NoStop}%
\bibitem [{\citenamefont {Bottaro}\ \emph {et~al.}(2022{\natexlab{a}})\citenamefont {Bottaro}, \citenamefont {Buttazzo}, \citenamefont {Costa}, \citenamefont {Franceschini}, \citenamefont {Panci}, \citenamefont {Redigolo},\ and\ \citenamefont {Vittorio}}]{Bottaro:2021snn}%
  \BibitemOpen
  \bibfield  {author} {\bibinfo {author} {\bibfnamefont {S.}~\bibnamefont {Bottaro}}, \bibinfo {author} {\bibfnamefont {D.}~\bibnamefont {Buttazzo}}, \bibinfo {author} {\bibfnamefont {M.}~\bibnamefont {Costa}}, \bibinfo {author} {\bibfnamefont {R.}~\bibnamefont {Franceschini}}, \bibinfo {author} {\bibfnamefont {P.}~\bibnamefont {Panci}}, \bibinfo {author} {\bibfnamefont {D.}~\bibnamefont {Redigolo}}, \ and\ \bibinfo {author} {\bibfnamefont {L.}~\bibnamefont {Vittorio}},\ }\href {\doibase 10.1140/epjc/s10052-021-09917-9} {\bibfield  {journal} {\bibinfo  {journal} {Eur. Phys. J. C}\ }\textbf {\bibinfo {volume} {82}},\ \bibinfo {pages} {31} (\bibinfo {year} {2022}{\natexlab{a}})},\ \Eprint {http://arxiv.org/abs/2107.09688} {arXiv:2107.09688 [hep-ph]} \BibitemShut {NoStop}%
\bibitem [{\citenamefont {{ATLAS collaboration}}(2018)}]{ATLAS:2018jjf}%
  \BibitemOpen
  \bibfield  {author} {\bibinfo {author} {\bibnamefont {{ATLAS collaboration}}},\ }\href@noop {} {\enquote {\bibinfo {title} {{ATLAS sensitivity to winos and higgsinos with a highly compressed mass spectrum at the HL-LHC}},}\ } (\bibinfo {year} {2018})\BibitemShut {NoStop}%
\bibitem [{\citenamefont {Bottaro}\ \emph {et~al.}(2022{\natexlab{b}})\citenamefont {Bottaro}, \citenamefont {Buttazzo}, \citenamefont {Costa}, \citenamefont {Franceschini}, \citenamefont {Panci}, \citenamefont {Redigolo},\ and\ \citenamefont {Vittorio}}]{Bottaro:2022one}%
  \BibitemOpen
  \bibfield  {author} {\bibinfo {author} {\bibfnamefont {S.}~\bibnamefont {Bottaro}}, \bibinfo {author} {\bibfnamefont {D.}~\bibnamefont {Buttazzo}}, \bibinfo {author} {\bibfnamefont {M.}~\bibnamefont {Costa}}, \bibinfo {author} {\bibfnamefont {R.}~\bibnamefont {Franceschini}}, \bibinfo {author} {\bibfnamefont {P.}~\bibnamefont {Panci}}, \bibinfo {author} {\bibfnamefont {D.}~\bibnamefont {Redigolo}}, \ and\ \bibinfo {author} {\bibfnamefont {L.}~\bibnamefont {Vittorio}},\ }\href {\doibase 10.1140/epjc/s10052-022-10918-5} {\bibfield  {journal} {\bibinfo  {journal} {Eur. Phys. J. C}\ }\textbf {\bibinfo {volume} {82}},\ \bibinfo {pages} {992} (\bibinfo {year} {2022}{\natexlab{b}})},\ \Eprint {http://arxiv.org/abs/2205.04486} {arXiv:2205.04486 [hep-ph]} \BibitemShut {NoStop}%
\bibitem [{\citenamefont {Capdevilla}\ \emph {et~al.}(2021)\citenamefont {Capdevilla}, \citenamefont {Meloni}, \citenamefont {Simoniello},\ and\ \citenamefont {Zurita}}]{Capdevilla:2021fmj}%
  \BibitemOpen
  \bibfield  {author} {\bibinfo {author} {\bibfnamefont {R.}~\bibnamefont {Capdevilla}}, \bibinfo {author} {\bibfnamefont {F.}~\bibnamefont {Meloni}}, \bibinfo {author} {\bibfnamefont {R.}~\bibnamefont {Simoniello}}, \ and\ \bibinfo {author} {\bibfnamefont {J.}~\bibnamefont {Zurita}},\ }\href {\doibase 10.1007/JHEP06(2021)133} {\bibfield  {journal} {\bibinfo  {journal} {JHEP}\ }\textbf {\bibinfo {volume} {06}},\ \bibinfo {pages} {133} (\bibinfo {year} {2021})},\ \Eprint {http://arxiv.org/abs/2102.11292} {arXiv:2102.11292 [hep-ph]} \BibitemShut {NoStop}%
\bibitem [{\citenamefont {Fukuda}\ \emph {et~al.}(2023)\citenamefont {Fukuda}, \citenamefont {Moroi}, \citenamefont {Niki},\ and\ \citenamefont {Wei}}]{Fukuda:2023yui}%
  \BibitemOpen
  \bibfield  {author} {\bibinfo {author} {\bibfnamefont {H.}~\bibnamefont {Fukuda}}, \bibinfo {author} {\bibfnamefont {T.}~\bibnamefont {Moroi}}, \bibinfo {author} {\bibfnamefont {A.}~\bibnamefont {Niki}}, \ and\ \bibinfo {author} {\bibfnamefont {S.-F.}\ \bibnamefont {Wei}},\ }\href@noop {} {\  (\bibinfo {year} {2023})},\ \Eprint {http://arxiv.org/abs/2310.07162} {arXiv:2310.07162 [hep-ph]} \BibitemShut {NoStop}%
\end{thebibliography}%

\end{document}